# Physical properties of a new cuprate superconductor $Pr_2Ba_4Cu_7O_{15-\delta}$

Akiyuki Matsushita[a], Kazunori Fukuda[b,d], Yuh Yamada[c,d], Fumihiro Ishikawa[b,d], Syun Sekiya[b], Masato Hedo[e], Takashi Naka[f]

[a]National Institute for Materials Science, 1-2-1 Sengen, Ibaraki 305-0047, Japan

[b]Graduate School of Science and Technology, Niigata University, Niigata 950-2181, Japan

[c]Department of Physics, Niigata University, Niigata 950-2181, Japan

[d]Center for Transdisciplinary Research, Niigata University, Niigata 950-2181, Japan

[e]The Institute for Solid State Physics, The University of Tokyo, Tokyo 277-8581, Japan

[f]Institute of Multidisciplinary Research for Advanced Materials, Tohoku Univ., Sendai 908-8577, Japan

**Abstract**

We present studies of the thermal, magnetic and electrical transport properties of reduced polycrystalline $Pr_2Ba_4Cu_7O_{15-\delta}$ (Pr247) showing a superconducting transition at $T_c$ = 10 - 16 K and compare them with those of as-sintered non-superconducting Pr247. The electrical resistivity in the normal state exhibited $T^2$ dependence up to approximately 150 K. A clear specific heat anomaly was observed at $T_c$ for Pr247 reduced in a vacuum for 24 hrs, proving the bulk nature of the superconducting state. By the reduction treatment, the magnetic ordering temperature $T_N$ of Pr moments decreased from 16 to 11 K, and the entropy associated with the ordering increased, while the effective paramagnetic moments obtained from the DC magnetic susceptibility varied from 2.72 to 3.13 $\mu_B$. The sign of Hall coefficient changed from positive to negative with decreasing temperature in the normal state of a superconducting Pr247, while that of as-sintered one was positive down to 5 K. The electrical resistivity under high magnetic fields was found to exhibit $T^\alpha$ dependence ($\alpha$ = 0.08 - 0.4) at low temperatures. A possibility of superconductivity in the so-called CuO double chains is discussed.

# 1. Introduction

Since the discovery of high-$T_c$ cuprate superconductors, extensive studies on both superconducting and normal states have been made on the basis of the physical properties of the $CuO_2$ planes. The drastic suppression of superconductivity by Pr substitution has been the object of study for a long time; it is now ascribed to the localization of carriers in the $CuO_2$ planes caused by the hybridization of Pr-4f and O-2p in the $CuO_2$ planes [1]. After the discovery of high-$T_c$ superconductivity in a reduced $Pr_2Ba_4Cu_7O_{15-\delta}$ (Pr247) [2], the Pr-based cuprate has attracted a significant amount of interest [3-5] because the superconductivity in the $CuO_2$ planes is considered to be suppressed in this compound owing to the Pr-4f – O-2p hybridization as described above, and, therefore, the possibility of superconductivity in a different structural unit, i.e., the CuO double chains, has been proposed. Pr247 contains three kinds of Cu-O structural units in the crystal structure, i.e., the $CuO_2$ plane, the CuO single chain, and the CuO double chain [2]. This structural feature gives Pr247 a unique property. In the CuO single chain, oxygen atoms are easily absorbed or desorbed; thus, the charge carrier density of the compound can be controlled utilizing this characteristic property. On the other hand, the CuO double chain is so rigid that it is considered to be free of oxygen deficiencies. As a result, a special situation is realized in Pr247; the carrier density of the CuO double chains in addition to that of the $CuO_2$ planes might be tuned by reduction or oxidization processes. Yamada et al. [3] studied the effect of the reduction treatment for the superconducting transition temperature $T_c$, the structural parameters, and the transport properties and found that the sign of the major carriers of the superconducting Pr247 is negative, in contrast to the sign in conventional high-$T_c$ cuprate superconductors. Watanabe et al. [4] reported that the $CuO_2$ planes were considered to be magnetically ordered in the superconducting Pr247 from the results of NQR measurements and, therefore, discarded the possibility of superconductivity in the $CuO_2$ planes.

It is not easy to obtain a high-quality superconducting sample in Pr247; the specimen undergoes a reduction treatment after an oxidization procedure under high pressure oxygen atmosphere. Since the first report on the superconductivity of Pr247 [2], we have been involved in improving the sample quality and, recently, obtained a sample in which a specific anomaly associated with the superconducting transition was observed for the first time. Although the sample still includes a small amount of the second phase at present, some physical properties such as transport properties are expected to represent intrinsic ones since the second phase is an insulator. In this paper, we report some physical properties of superconducting Pr247 and show that they are considerably different from those of conventional high-$T_c$ compounds.

## 2. Experimental

Polycrystalline Pr247 compounds were synthesized using a high-oxygen-pressure technique. The precursor was prepared using a polymerized complex method of $Pr_6O_{11}$, CuO, and $Ba(NO_3)_2$. This mixture was decomposed at 1073-1153 K in air for 24 hr. The resultant powder was pressed into a pellet and then reacted in pure oxygen gas of 5 atm at 973 K for 18 hr (as-sintered sample). The reduced sample was prepared by post-annealing in a vacuum at 673 K. In the following, reduced samples are denoted by their reduction time, such as a 24 hr-reduced sample. Figure 1 shows the X-ray diffraction profiles for the as-sintered sample and the 24 hr-reduced one. The samples were confirmed to be the Y247 structure containing a small amount of the $BaCuO_2$ phase. The detailed structural analysis based on the powder synchrotron X-ray diffraction performed at the Spring-8 beamline BL02B2 is shown in ref. 3. From the weight change of the sample before and after the reduction treatment, the oxygen deficiency $\delta$ was estimated to be 0.5 for the 24hr-reduced sample, which showed the largest superconducting volume fraction in a magnetic measurement. In Fig.2, the variation of magnetic susceptibility associated with the superconducting transition is shown for the 24hr-reduced sample, indicating the bulk nature of the superconductivity. The amount of the second phase $BaCuO_{2+x}$ was estimated to be approximately 9 % for all samples using Rietvelt analysis. A certain amount of mixing between the Pr and Ba atoms was observed [3].

The magnetic susceptibility was measured using a SQUID magnetometer (MPMS of Quantum Design Co.). The electrical resistivity and Hall effect were measured using conventional four-probe and five-probe methods, respectively. The specific heat was measured between 2 and 100 K with a thermal relaxation technique. These three measurements were carried out using the PPMS of Quantum Design Co. The $BaCuO_{2+x}$ phase exhibits considerably large specific heat at low temperatures [6], but the error coming from this phase in specific heat was estimated to be approximately 3 % above 15 K.

## 3. Results and discussion

### 3.1. Electrical resistivity

Figure 3 shows the electrical resistivity as a function of the temperature for a 9 hr-reduced sample that exhibited a superconducting transition at 15 K. In the normal state, the electrical resistivity clearly shows $T^2$ dependence up to approximately 150 K. To determine the value of the exponent of temperature, we carried out a least-squares fit for the data of 6, 9, and 24hr-reduced samples in the temperature range of 20 - 100 K. The values of the exponent of temperature were obtained to be 2.0, 2.0, and 1.9 for 6hr-, 9hr-, and 24hr-reduced samples, respectively. In a 3hr-reduced sample, the $T^2$ dependence was also observed, but it was limited to a narrower temperature range. Thus, the $T^2$ dependence was widely observed in the Pr247 samples that underwent different reduction treatments and is considered to be one of the most distinctive characters of superconducting Pr247. This behavior is in clear contrast to the famous $T$-proportional dependence widely observed in conventional high-$T_c$ cuprates. We believe that this $T^2$ dependence originates from the conduction in the so-called CuO double chains for the reasons described in the following. First, in a measurement for PrBa$_2$Cu$_4$O$_8$ (Pr124) single crystals, the conduction through the CuO double chains was experimentally shown to be metallic [7], while the CuO$_2$ planes were supposed to be insulating. Consequently, the metallic resistivity in Fig. 3 is expected to originate from that in the CuO double chains. Second, the electrical resistivity through the CuO double chains was indicated to display $T^2$ dependence in a wide temperature range in YBa$_2$Cu$_4$O$_8$ (Y124) [8]. Interestingly, some other quasi-one-dimensional materials were also reported to show $T^2$ dependence up to about room temperature, for example, the CuO single chain of YBa$_2$Cu$_3$O$_{6.9}$ [9] and polymeric sulfur nitride (SN)$_x$ [10]. Thus, it is likely that the $T^2$ behaviors are observed in general up to much higher temperature in quasi-one-dimensional materials than in three dimensional materials such as the heavy fermion materials, in which the $T^2$ dependence is usually a phenomenon below a few K. Basically, $T^2$ dependence has been ascribed to electron-electron scatterings, which are, in conventional situations, overwhelmed by phonon scattering at high temperatures. We do not have a convincing reason at present to explain why the electron-electron scattering is so prominent relative to the phonon scattering in this material, but one possible explanation is that the flat Fermi surface, which may be expected in a quasi-one-dimensional material, restricts the allowed phonon scattering, leading to its suppression. Thus, the $T^2$ dependence of the electrical resistivity in Pr247 is assumed to be an interesting aspect of the conduction in the CuO double chains.

## 3.2. Hall coefficient

The charge carriers of most high-$T_c$ cuprate superconductors are holes. The hole density can be increased by an oxidation treatment, which leads to a rise of $T_c$ for under-doped materials, although the oxidation treatment is not always applicable to all high-$T_c$ cuprates. In contrast, Pr247 has to be reduced to give rise to superconductivity. Therefore, it is expected that the charge carriers are electrons in a superconducting Pr247.

Figure 4 shows the effect of the reduction treatment on the Hall coefficient. In an as-sintered sample, the Hall coefficient was basically positive in the whole temperature region, which is consistent with the expectation that the charge carriers of Pr247 should be holes, as stated later. However, the behavior of the Hall coefficient dramatically changed after only a 3hr reduction treatment; the Hall coefficient was positive around room temperature but showed a rapid turn to negative below 200 K. This behavior clearly shows that a contribution of negative charge carriers appeared after reduction treatment, and holes and electrons coexisted in the reduced Pr247. In the 24hr-reduced sample, the temperature dependence is quite similar to that of the 3hr-reduced sample, but the values showed a positive shift in the whole temperature region. Here, we assume that there are two kinds of carriers, i.e., holes and electrons, which can be characterized by their densities and mobilities: hole density $p$, electron density $n$, hole mobility $\mu_h$, and electron mobility $\mu_e$. Then, the Hall coefficient $R_H$ can be written as,

$$R_H = \frac{1}{ec} \cdot \frac{p - nb^2}{(p + nb)^2} \qquad b = \frac{\mu_e}{\mu_h}.$$

Using this expression, we can easily show that the absolute value of $R_H$ would decrease with increasing $n$ if $n$ is larger than a certain value. Therefore, the positive shift of $R_H$ in the negative $R_H$ region can be basically interpreted as the result of an increase in electron density owing to the longer reduction treatment. However, the positive shift of $R_H$ in the positive $R_H$ region can not be simply explained.

Thus, the behavior of the Hall effect in the reduced Pr247 is not thoroughly understood at present, but the point is that negative charge carriers were really introduced by the reduction treatments. In general, the sign of charge carriers in high-$T_c$ cuprates strongly depends on the crystal structure. The apical oxygen, an oxygen atom adjacent to the $CuO_2$ plane, plays an especially decisive role; the charge carriers are holes if the structure has apical oxygen. For example, $(Nd,Ce)_2CuO_4$ is a cuprate superconductor in which the major charge carriers are electrons. The crystal structure of (Nd,Ce)$_2$CuO$_4$ is called a T' structure, which lacks the apical oxygen. Pr247 is a Pr-analog of the Y$_2$Ba$_4$Cu$_7$O$_{15-\delta}$ (Y247) superconductor in the YBa$_2$Cu$_{3+0.5n}$O$_{7+0.5n}$ (n = 0, 1, 2) series, which is a typical hole-type high-$T_c$ superconductor. Therefore, Pr247 should be a hole-type superconductor from its structure as far as we suppose the conduction in the CuO$_2$ planes. In fact, the major carriers in as-sintered Pr247 are holes, as shown in Fig. 4. Thus, it is unlikely that electrons are introduced into the CuO$_2$ planes in Pr247. Consequently, we assume that electrons are doped in the CuO double chains.

*3.3. Magnetic susceptibility*

The investigation on the electronic states of Pr which is located between the CuO$_2$ planes is the key to an understanding of the state of the CuO$_2$ planes. In fact, the theory of the hole transfer from CuO$_2$ planes to the hybridization states between Pr-4f and O-2p [1] is widely accepted. In Fig. 5, the DC magnetic susceptibility $\chi$ above $T_N$ is shown in $1/(\chi-\chi_0)$ vs. $T$ plot for the as-sintered and the 24hr-reduced samples, where $\chi_0$ is a constant. It is clear that both data can be well fitted with the modified Curie-Weiss law $\chi - \chi_0 = C_m/(T-\theta)$ up to room temperature, where $C_m$ and $\theta$ are constants. The effective moments for Pr were obtained from the Curie constant $C_m$ to be 2.7 and 3.1 $\mu_B$ for the as-sintered sample and the 24hr-reduced one, respectively. It should be noted that these values include the contribution of BaCuO$_{2+x}$ which is estimated to be about 10%. However, the effect of reduction on the magnetism of BaCuO$_{2+x}$ is not so large [11] to change the following discussion. We may say that the values of Pr magnetic moments lie between the free-ion values of Pr$^{4+}$ (2.54 $\mu_B$) and Pr$^{3+}$ (3.58 $\mu_B$) and that the valence of Pr approaches to 3+ with the reduction treatment. This behavior can be explained quite naturally as due to the reduction of Pr ions with the reduction treatment. Although the problem of the valence of Pr is more complicated [12], the interpretation that the Pr ions were reduced with the reduction treatment is supported by the specific heat measurements as mentioned later.

*3.4. Specific heat*

The specific heat anomaly associated with the superconducting transition is shown in Fig. 6 for a 24hr-reduced sample. The steep rise in $C/T$ below 15 K is owing to the magnetic ordering of Pr. To better appreciate the details of the specific heat anomaly, we fitted the total specific heat far from $T_c$ with a polynomial and subtracted this function from the data around $T_c$. The result is shown in the inset of Fig. 6 and the specific heat jump $\Delta C/T_c$ was evaluated to be about 8 mJmol$^{-1}$K$^{-2}$. This value is fairly small compared to other high-$T_c$ materials, but this value is supposed to increase as the sample quality improves because the synthesis of a superconducting Pr247 is not easy and the sample quality is still improving. However, the occurrence of an anomaly is, in principle, a proof of the bulk nature of the superconducting state.

In the previous section the possibility that the valence of Pr approached to 3+ after the reduction treatment has been pointed out. Here, we examine a specific heat anomaly associated with the magnetic ordering of Pr moments. Figure 7 shows the effects of the reduction treatment on the specific heat around the anomaly associated with the magnetic ordering in $C/T$ vs. $T^2$ plot. This compound, as has been reported, contains an appreciable amount of BaCuO$_{2+x}$, which may exhibit large specific heat depending on the value of x [6]. In addition, a certain amount of Pr substitutes for the Ba sites. Therefore, we are unable to provide a quantitative discussion, but some trends are obvious. First, the magnetic ordering temperature $T_N$, which was 16 K in an as-sintered sample, decreased to 11 K after a 24hr reduction treatment. Secondly, the entropy associated with the magnetic ordering increased with the reduction treatment. Thirdly, the $T$-proportional term of the specific heat showed an increase after the treatment. To confirm these effects of the reduction treatment, we fitted the C/T data with a equation of $C/T = \gamma + \beta T^2$ in the temperature range of 20 to 40 K, as shown in Fig. 7, where $\gamma$ and $\beta$ are constants. We then subtracted this function from the data to roughly estimate the entropy associated with the peak of the magnetic ordering. The entropy thus obtained was 1.7 and 3.5 J/K per mol Pr for the as-sintered and 24hr-reduced samples, respectively. These values are far smaller than those reported for Pr123 [13] and Pr124 [14]. Probably, such a straightforward analysis is inappropriate for obtaining a correct value. However, the situation is complex, and definite information is lacking; therefore, a detailed discussion on this matter would be too time-consuming at present. Further investigation will be necessary to clarify this point. The $\gamma$ values were 0.32 JK$^{-2}$(mol Pr)$^{-1}$ for the as-sintered sample and 0.35 for the 24hr-reduced sample, where "mol Pr" means a specific heat per 0.5 mol Pr$_2$Ba$_4$Cu$_7$O$_{15-\delta}$. The absolute value of $\gamma$ includes a certain amount of errors owing to the contribution of the second-phase BaCuO$_{2+x}$, but we may say that the $\gamma$ value of Pr247 increased after the reduction treatment since the fraction of the BaCuO$_{2+x}$ phase is almost same in both samples and, therefore, the contribution to the specific heat is also expected to be the same. The Debye temperatures obtained from the $\beta$ values were approximately 340 K for both samples. These effects brought about by the reduction treatment could be interpreted as follows. The reduction treatment is considered to fill the holes in the CuO$_2$ planes and, therefore, to fill the holes in the hybridization states between Pr-4f and O-2p, leading to a decrease in the valence of the Pr atom; in other words, Pr ions were reduced to approach a valence of 3+. As a result, the fraction of Pr$^{3+}$ increased, and the entropy associated with Pr ordering became larger. The decrease of the hole density in the CuO$_2$ planes, in turn, may weaken the magnetic interaction between Pr moments, leading to the decrease of $T_N$. On the other hand, the hole-filling effect in the CuO$_2$ planes is unlikely to explain the increase of the $\gamma$ value. However, as noted above, if the reduction treatment injects electrons into the CuO double chains as well as the CuO$_2$ planes, the increase of the $\gamma$ value can be explained as an increase of the negative charge carriers in the CuO double chains.

It is known that many *f*-electron-based systems show a universal behavior, $A/\gamma^2 = 1.0*10^{-5}$ µΩcm(Kmol/mJ)$^2$, called the Kadowaki-Woods (KW) relation [15], where A is the coefficient of $T^2$ in the electrical resistivity, as shown in Fig. 3. It would be worthwhile to see whether or not the reduced Pr247 obeys the KW relation. For the 24hr-reduced sample, A and $\gamma$ were 0.41 ΩcmK$^{-2}$ and 350 mJK$^{-2}$(mol Pr)$^{-1}$, respectively. Thus, we obtain $A/\gamma = 0.33*10^{-5}$ Ωcm(Kmol/mJ)$^2$, which is slightly smaller than the value expected in the KW relation.

However, according to a recent study [16], the A/γ value may become smaller than the KW relation depending on the degeneracy of quasi-particles. Thus, the KW relation seems to be valid in the reduced Pr247. This is a somewhat unexpected result because the $T^2$ dependence is considered to be the property of the CuO double chains while the γ value obviously includes the contribution from the $CuO_2$ planes. We need a detailed study on a sample of better quality to clarify this problem.

*3.5. Electrical resistivity under high magnetic fields*

In the above sections, we have shown that the physical properties of the reduced Pr247 are very unique comparing with those of conventional high-$T_c$ materials and have indicated the possibility that superconductivity occurs in the CuO double chains. In this section, we show another characteristic property that was observed in all of the reduced Pr247 superconductors. Figure 8(a) shows the electrical resistivity of the 24hr-reduced sample under high magnetic fields and it is replotted as log $T$ vs. log $\rho$ in Fig. 8(b). As seen in the figure, at high magnetic fields, the electrical resistivity was found to show $T^\alpha$ dependence below approximately 10 K. This $T^\alpha$ dependence was observed in all other superconducting samples, and the values of $\alpha$ are listed in Fig. 8(b) together with the reduction time $t_R$ and $T_c$ for all samples. Interestingly, it is seen that the values of $\alpha$ exhibit a good correlation with $T_c$.

Next, we tried to determine the lower temperature limit for the $T^\alpha$ dependence. As the temperature decreased below 5 K, all samples except the 3hr-reduced sample showed a deviation from the $T^\alpha$ dependence at a certain temperature owing to superconductivity. In the 3hr-reduced sample, in which the superconducting state is considered to be the weakest against the magnetic field, the $T^\alpha$ dependence is observed at lower temperatures, as shown in Fig. 9. In the lower magnetic fields, the electrical resistivity deviates from the $T^\alpha$ dependence, as indicated by arrows, but, at 7 T, the $T^\alpha$ dependence was observed to be down to about 0.3 K. If much higher magnetic fields were available, the $T^\alpha$ dependence would be observed at lower temperatures in other samples as well. The origin of the $T^\alpha$ dependence is not clear at present. It might be, e.g., a superconducting fluctuation caused by the mixing of Pr and Ba ions. However, it is noteworthy that the $T^\alpha$ behavior is predicted in the Tomonaga-Luttinger-liquid theory [17,18]. This theory was developed for a one-dimensional system, and, therefore, we are assuming that the $T^\alpha$ behavior is related to the one-dimensionality of the CuO double chain.

## 6. Conclusion

We have presented the specific heat anomaly associated with the superconducting transition in reduced Pr247 superconductor for the first time. This is an evidence of the bulk nature of the superconducting state. Both specific heat and magnetic susceptibility indicate that Pr was reduced and approached to $Pr^{3+}$ after reduction treatment. We have also reported some unique transport properties. The first is the $T^2$ dependence in electrical resistivity, which was observed up to approximately 150 K. The second is the negative Hall coefficient. The charge carriers of Pr247 should be holes if they are in the $CuO_2$ planes from a structural point of view. Therefore, this fact strongly supports the supposition that the electrons are doped in the CuO double chains. The third is the $T^\alpha$ dependence of electrical resistivity under high magnetic fields. This behavior has been predicted by the Tomonaga-Luttinger-liquid theory, and further studies are expected to elucidate its origin and relationship with the CuO double chains. In this material, a certain amount of mixing between Pr and Ba is inevitable, and, as a result, the second-phase $BaCuO_{2+x}$ appears. Owing to these problems, the interpretations on some of the results have become difficult. We are now carrying out a synthesis of the superconducting single crystal of Pr247 to overcome these difficulties and to solve the remaining problems.

**Acknowlegement**

This research was partially supported by the Ministry of Education, Sciene, Sports and Culture, a Grant-in-Aid for Scientific Research and the Grant for Promotion of Niigata University Research Projects.

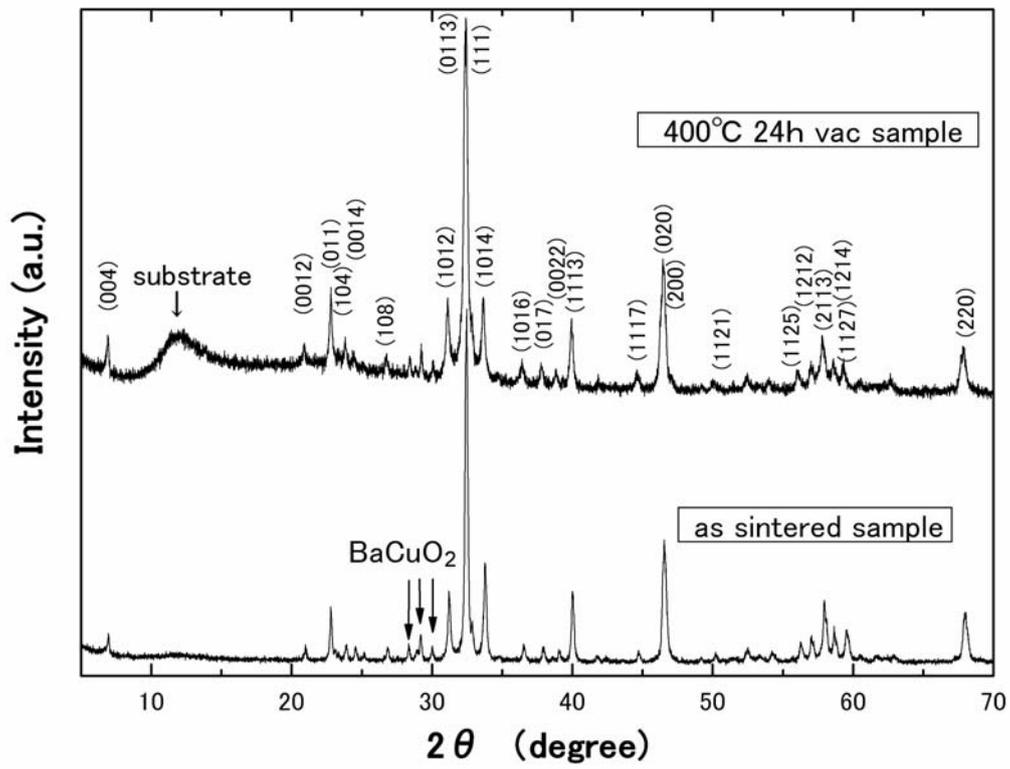

Fig. 1. Powder X-ray diffraction profiles for the as-sintered sample and the 24 hr-reduced sample.

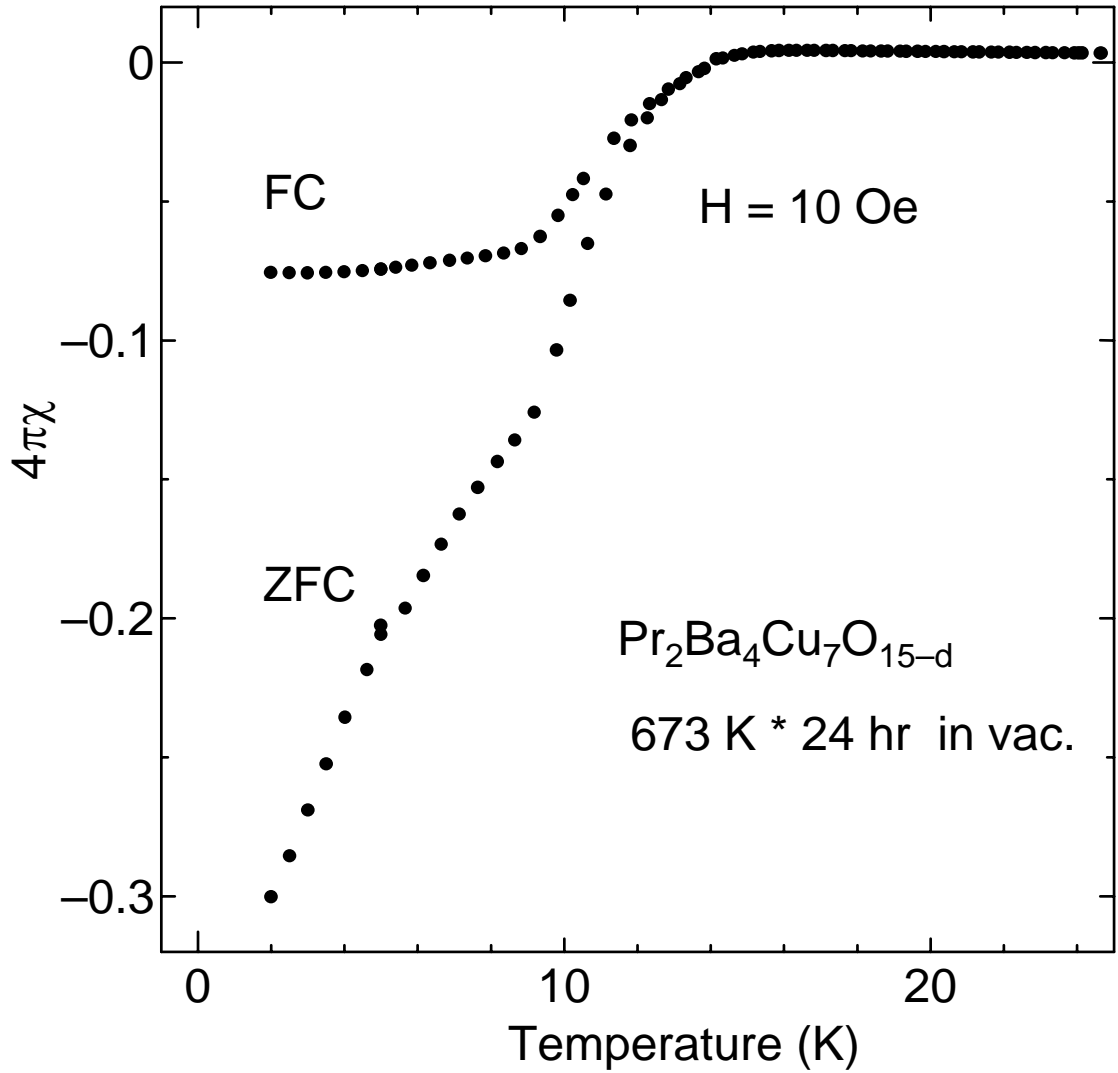

Fig. 2. Magnetic susceptibility $\chi$ multiplied by $4\pi$ for the 24 hr-reduced sample in the vicinity of $T_c$. FC and ZFC mean field-cool and zero-field-cool, respectively.

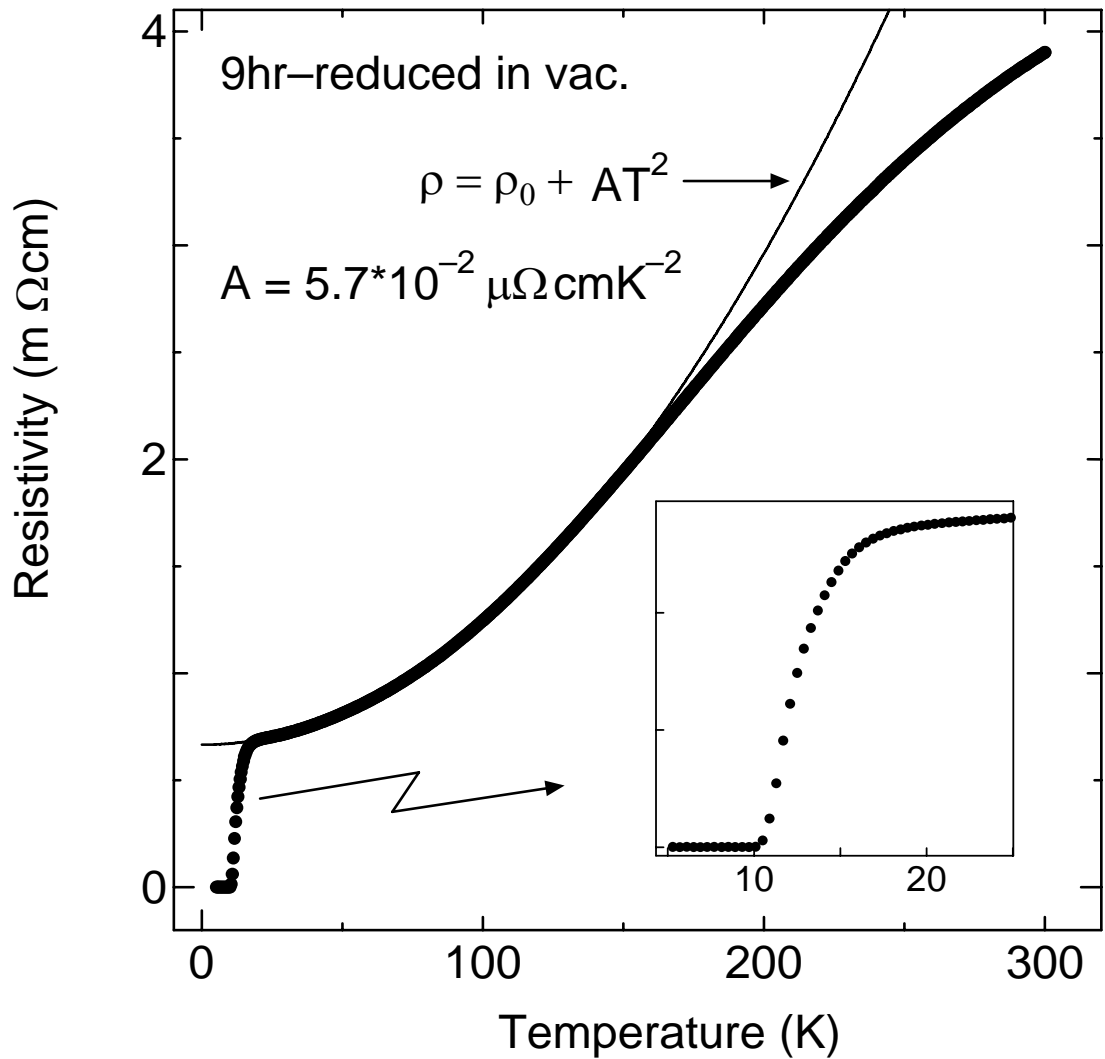

Fig. 3. Electrical resistivity of the 9hr-reduced $Pr_2Ba_4Cu_7O_{15-\delta}$ as a function of the temperature. The solid line is a fitting result with $\rho = \rho_0 + AT^2$, where $\rho$ is the electrical resistivity and $\rho_0$ and A are constants.

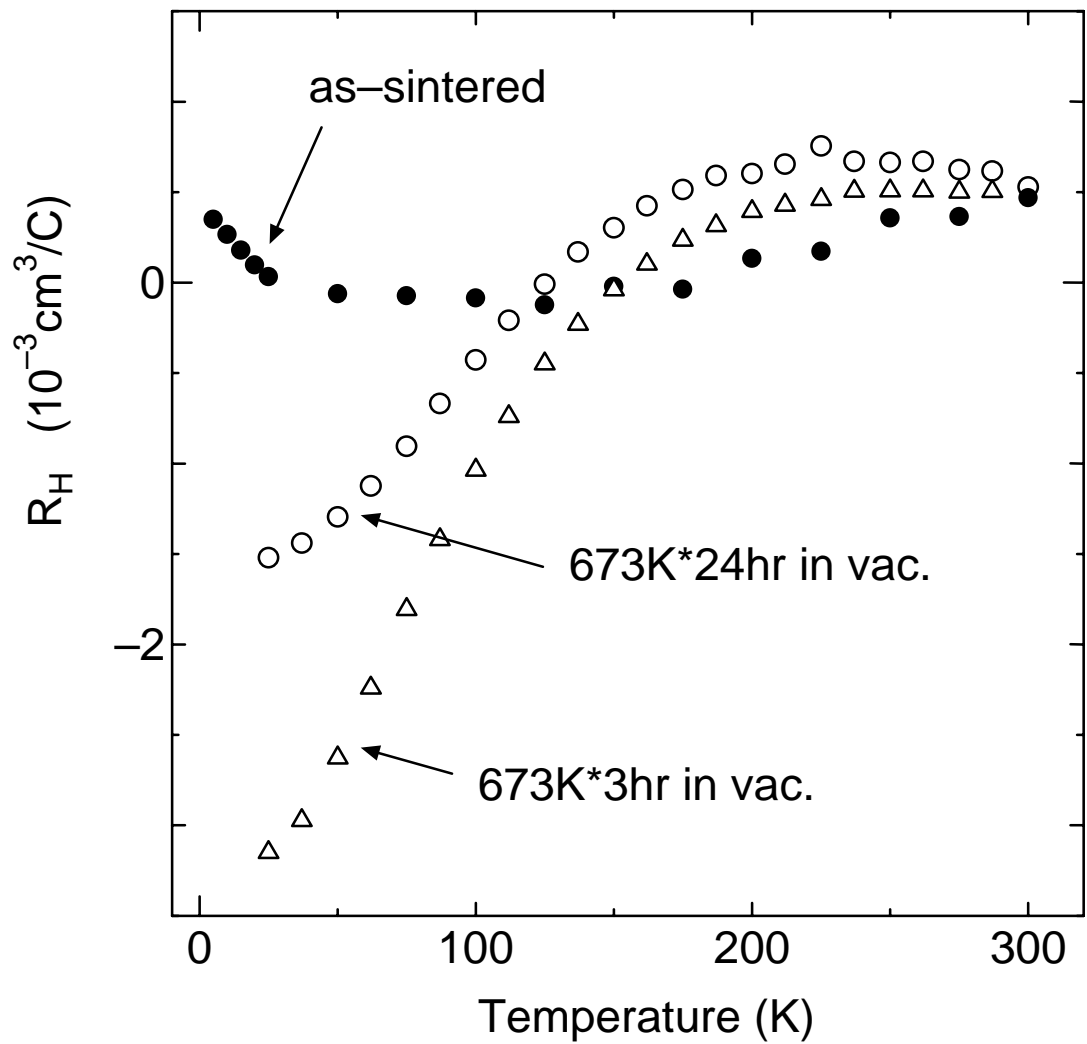

Fig. 4. Hall coefficient $R_H$ as a function of the temperature for the as-sintered sample, the 3hr-reduced sample, and the 24hr-reduced sample.

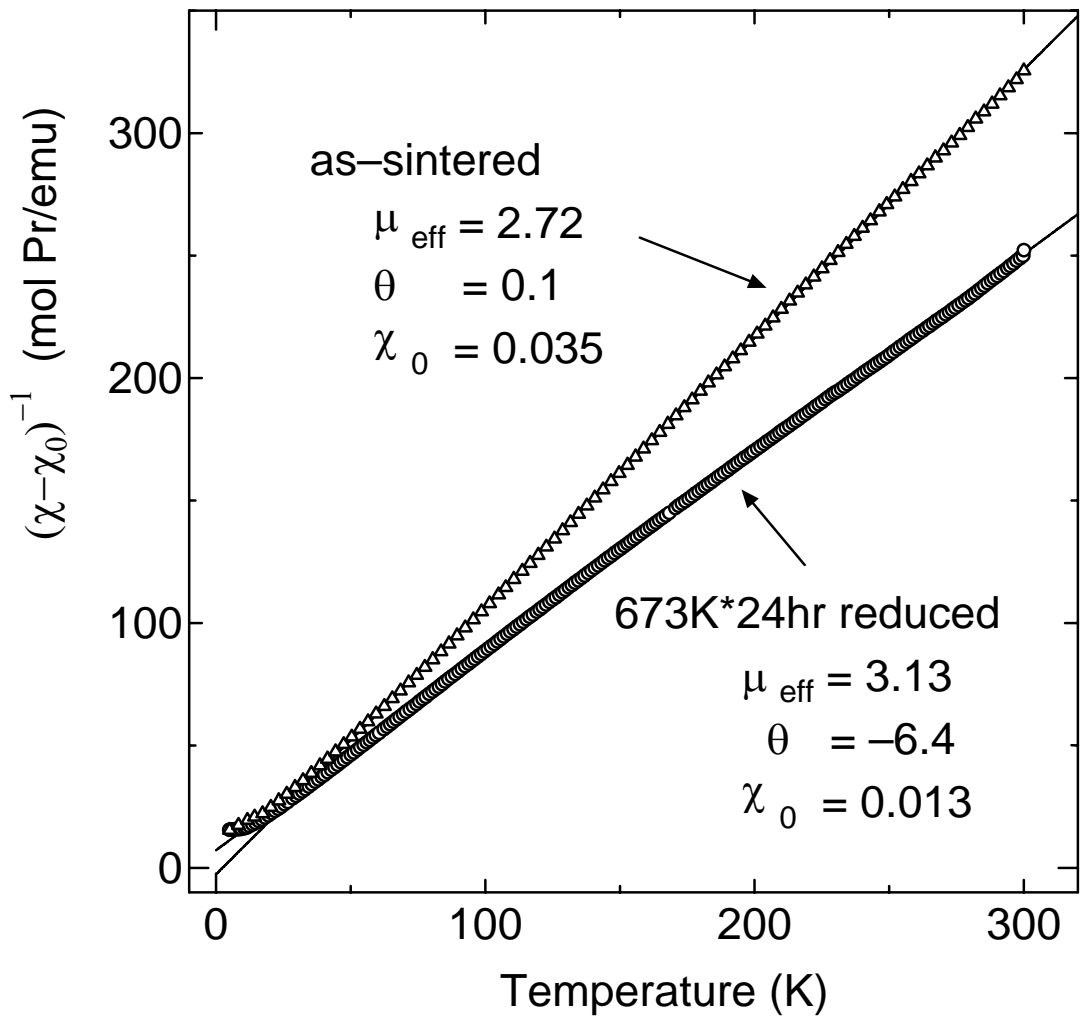

Fig. 5. Inverse magnetic susceptibility as a function of the temperature above $T_N$ for the as-sintered sample and the 24hr-reduced sample. $\chi_0$ is a fitting parameter. Details are given in the main text.

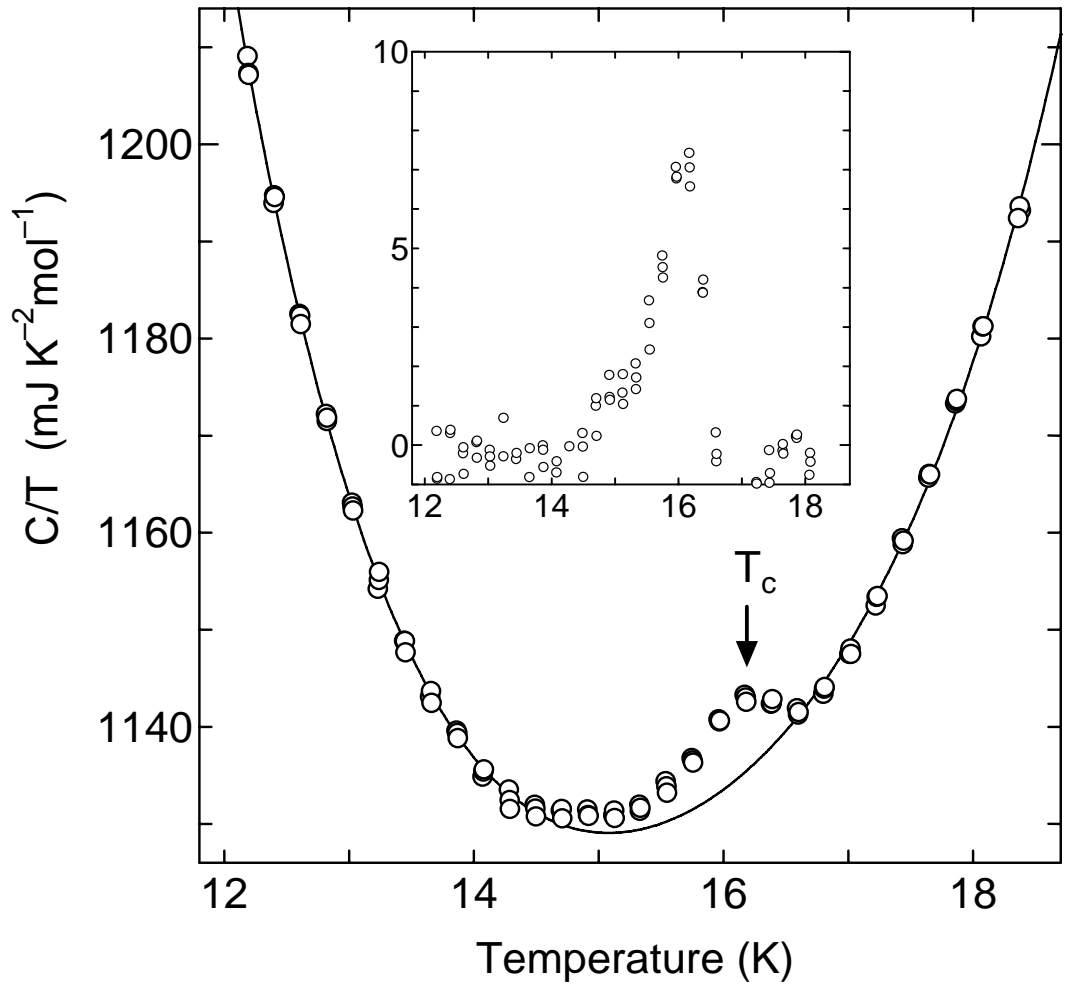

Fig. 6. Specific heat anomaly of the 24hr-reduced $Pr_2Ba_4Cu_7O_{15-\delta}$ in the vicinity of $T_c$. The inset shows the singular part of the specific heat.

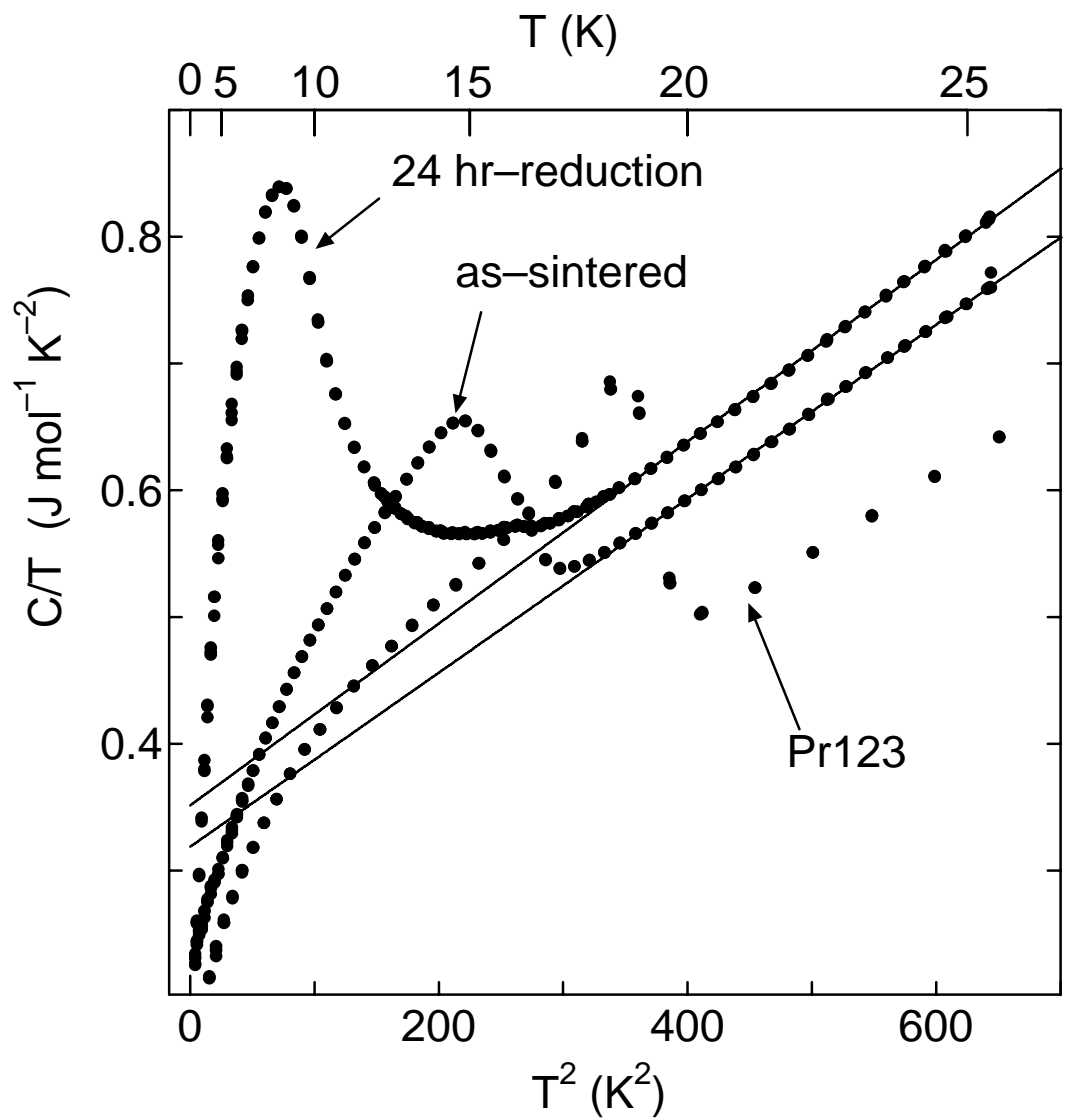

Fig. 7. Specific heat anomaly associated with the magnetic ordering of Pr moments for the as-sintered sample, the 24hr-reduced sample, and $PrBa_3Cu_3O_{7-\delta}$.

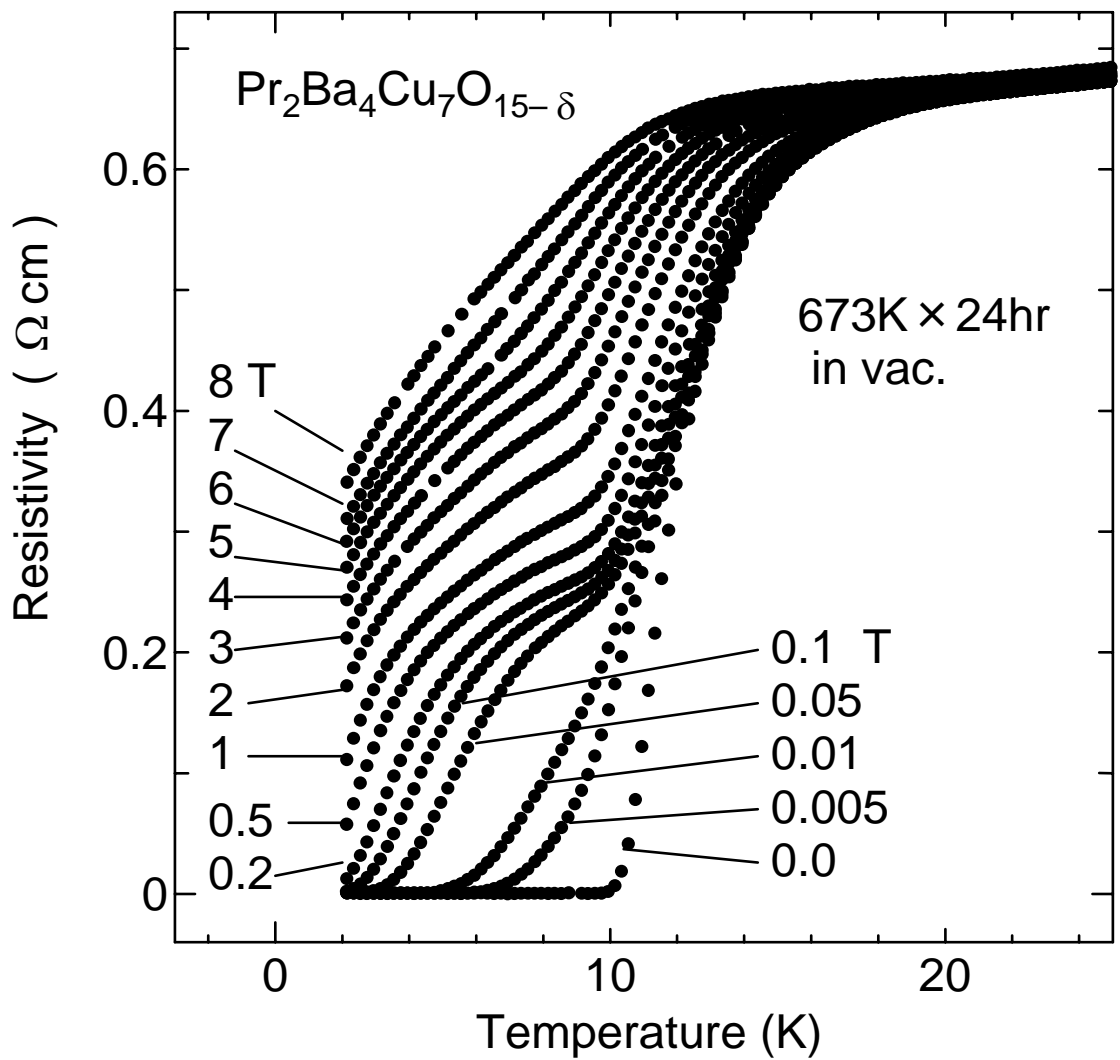

Fig. 8(a)

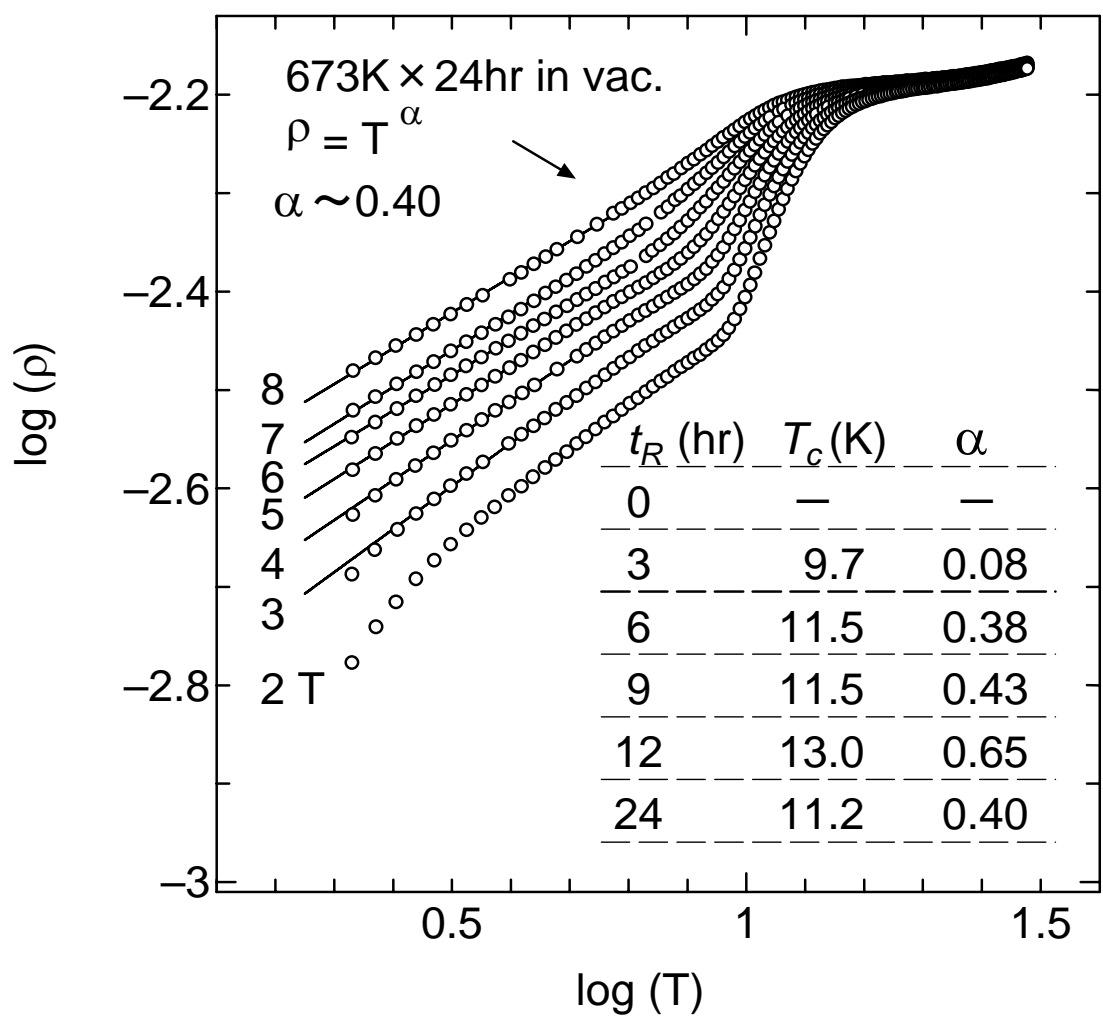

Fig. 8(b)

Fig. 8. Electrical resistivity of the 24hr-reduced sample under high magnetic fields plotted in (a) $\rho$ vs. $T$ and (b) $\log \rho$ vs. $\log T$. The values of $\alpha$, $T_c$, $t_R$ are also listed for all samples where $t_R$ is the reduction time.

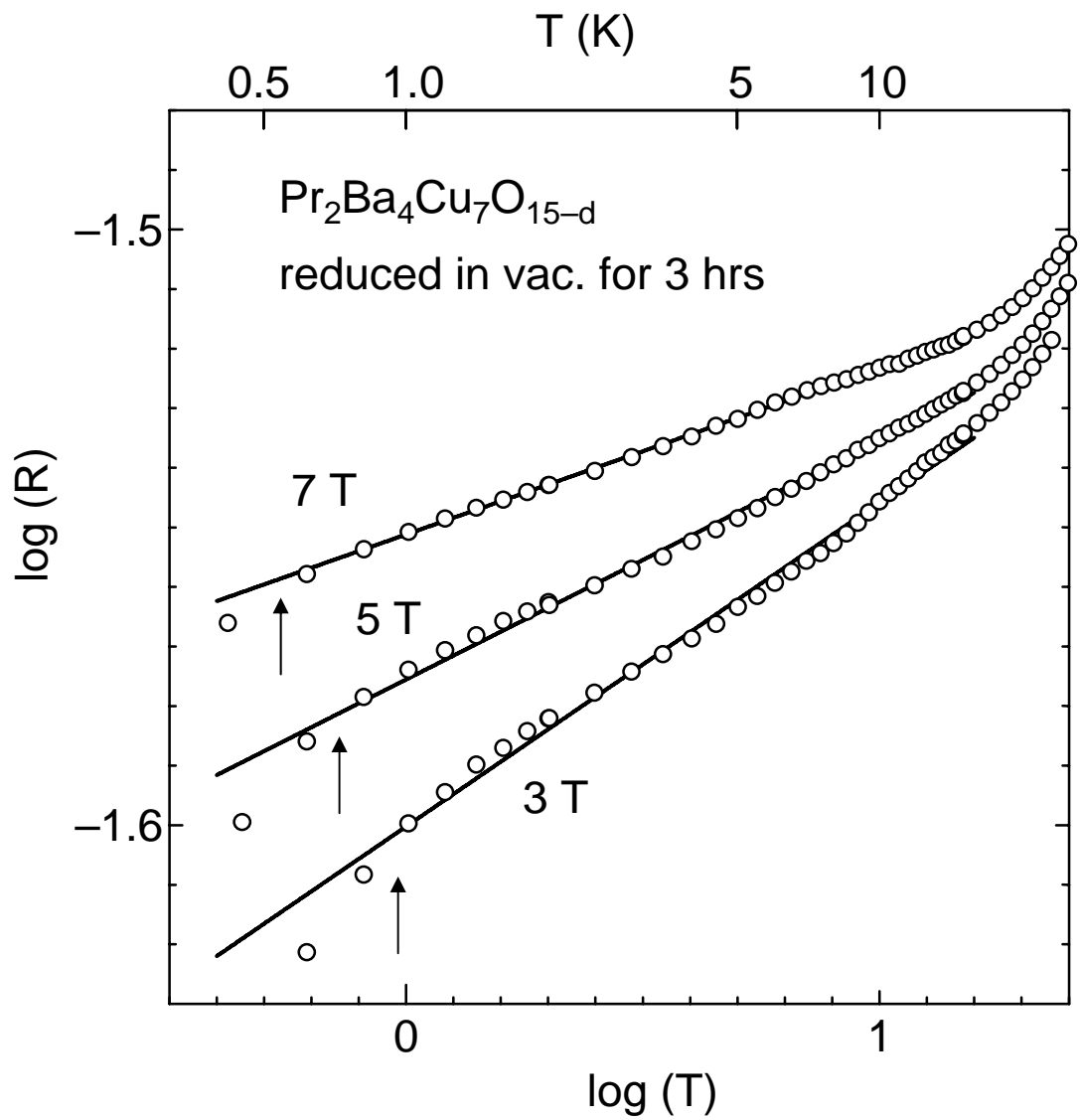

Fig. 9. Electrical resistivity of the 3hr-reduced sample under high magnetic fields measured down to 0.3 K.